# A comparison of two approaches for measuring interdisciplinary research output: the disciplinary diversity of authors vs the disciplinary diversity of the reference list[1]

Giovanni Abramo[1], Ciriaco Andrea D'Angelo[2], Lin Zhang[3,4*]

[1] Laboratory for Studies in Research Evaluation, Institute for System Analysis and Computer Science (IASI-CNR), National Research Council of Italy, Italy

[2] University of Rome "Tor Vergata" and Institute for System Analysis and Computer Science - National Research Council of Italy, Italy

[3] School of Information Management, Wuhan University, China

[4] Department of MSI, Centre for R&D Monitoring (ECOOM), KU Leuven, Belgium

**Abstract**

This study investigates the convergence of two bibliometric approaches to the measurement of interdisciplinary research: one based on analyzing disciplinary diversity in the reference list of publications, the other based on the disciplinary diversity of authors of publications. In particular we measure the variety, balance, disparity and integrated diversity index of, respectively, single-author, multi-author single-field, and multi-author multi-field publications. We find that, in general, the diversity of the reference list grows with the number of fields reflected in a paper's authors' list and, to a lesser extent, with the number of authors being equal the number of fields. Further, we find that when fields belonging to different disciplines are reflected in the authors' list, the disparity in the reference list is higher than in the case of fields belonging to the same discipline. However, this general tendency varies across disciplines, and noticeable exceptions are found at individual paper level.

**Keywords**

*Interdisciplinarity; research collaboration; co-authorship; bibliometrics; Italy*

# 1. Introduction

The possibility of scientific and social gain through interdisciplinary research (IDR) is of increasing interest to both academics and policymakers. Among many sources, the importance of this theme is attested by the data reported in the US NSF's 2016 Science and Engineering Indicators (National Science Board, 2016). Continuing the pattern of previous years, in 2014, around 2% of total federal U.S. spending for academic R&D in science and engineering was allocated to interdisciplinary or multidisciplinary research, not to a specific field. Additionally, within U.S. higher education, national survey data continues to show a tendency towards knowledge integration from multiple disciplines. Between 2004 and 2013, universities responding to the NSF's annual Higher Education Research and Development Survey reported steady growth in R&D spanning more than one field of science and engineering, and 40% of respondents to the NSF's Survey of Earned Doctorates[2] in 2013 reported two or more dissertation research fields, up from 24% in 2001.

Although present research policies often implicitly assume that IDR can be readily identified and tracked, this is far from true. Providing policymakers with measures and analyses that capture the intensity of IDR and knowledge integration is a scientific task of high practical importance, yet it is fraught with difficulties – see Wagner et al. (2011) and Rousseau, Zhang, & Hu (2018) for a review. In this work, we focus on the issues associated with measuring IDR. More precisely, we investigate the convergence of two bibliometric approaches to measurement: one based on analyzing disciplinary diversity in the reference lists of publications (Porter, Cohen, Roessner, & Perreault, 2007; Rafols & Meyer, 2010; Wang, Thijs, & Glänzel, 2015; Mugabushaka, Kyriakou, & Papazoglou, 2016; Zhang, Rousseau, & Glänzel, 2016), referred to as the reference list method in the following; the other based on the disciplinary diversity of a publication's authors (Schummer, 2004; Abramo, D'Angelo, & Di Costa, 2012), referred to as the authors method in the following. Measuring IDR has important benefits: among others, learning more about the collaboration behaviour of scientists, informing policies and initiatives aimed at fostering IDR, as well as monitoring its trends to assess the efficacy of policies.

The paper is organized as follows. After an overview of the literature on the subject, in Section 3 we present the field of observation and the way we apply the two methods to measure IDR: the authors method and the reference list method. In section 4 we illustrate the results of the analysis, and in Section 5 we draw our conclusions.

# 2. Literature review

Despite a growing attention to IDR by many scholars, there are still challenges on various fronts. Among them, developing a conceptual and practical definition of IDR, and indicators and methods to measure IDR (Huutoniemi, Klein, Bruun, & Hukkinen, 2010).

IDR can mean different things to different people. According to the Committee on the Science of Team Science et al. (2005), interdisciplinary research is: *"A mode of research by teams or individuals that integrates information, data, techniques, tools, perspectives, concepts and/or theories from two or more disciplines or bodies of specialized knowledge to advance fundamental understanding or to solve problems whose solutions are beyond the scope of a single discipline or area of research practice." In this definition, the key concept is knowledge integration. The more an article, or any other item under investigation, integrates sources from different disciplines, the more it is interdisciplinary. The sources could be information, data, techniques, tools, perspectives, concepts, etc.

---

[2] https://www.nsf.gov/statistics/srvydoctorates/#tabs-1. Last accessed 30 July, 2018.

In general, the literature classifies research activity involving experts of different disciplines as belonging to three principal categories: multidisciplinary, interdisciplinary, and transdisciplinary research (OECD, 1998). Stokols et al. (2003) provide a brief and precise distinction, as follows. "Multidisciplinary" research occurs when researchers from different disciplines work independently and sequentially, each from his or her own discipline-specific perspective, to address a common problem. In "interdisciplinary" research, researchers work jointly, but from the perspective of each of their respective disciplines to address a common problem. In "transdisciplinary" research, researchers work jointly to develop and use a shared conceptual framework that draws discipline-specific theories, concepts, and methods together to address a common problem. Choi and Pak (2006) contrast the different definitions of multidisciplinary, interdisciplinary, and transdisciplinary research in the literature. They find that the three terms are used to refer to a continuum of increasing levels of involvement by multiple disciplines. Multidisciplinarity sits at the base, where different disciplines work on the same problem in parallel, or sequentially, to move beyond the confines of their own field. Interdisciplinarity follows, where each discipline interacts reciprocally. Reaching this level requires a "loosening" of the disciplinary confines to generate new methodologies, knowledge, or even new shared disciplines. Finally, at the transdisciplinary level, each discipline transcends its traditional confines and examines the dynamics of entire systems from a holistic point of view. Although the distinctions between each of the above terms are valuable, evidence of the continuum found in empirical studies can often make it difficult to distinguish which is which (Rafols and Meyer 2010). According to a review by Klein (2008), each of these three types of "disciplinarity" is also characterized by a particular type of "knowledge integration", meaning a particular mode of merging theories and concepts, techniques and tools, or information and data from various fields of knowledge. In this paper, we use the term interdisciplinary (interdisciplinarity) in a more general sense to encompass multi-, trans-disciplinary research on the individual paper level.

However subtle and sophisticated these distinctions between the different modes of integrating knowledge might be, in the end, one has to face the reality of measuring such "knowledge integration" and the challenges they presents. It seems not recommendable to define a unique and absolute measure of IDR. Hence, scholars have developed a variety of proxy indicators, each one delivering different insights about the interdisciplinary nature of the research under study. It is, therefore, unsurprising that these different indicators sometimes deliver inconsistent and even contradictory results. Adams, Loach, & Szomszor (2016) point out that it is essential to consider a framework for analysis that draws on multiple indicators rather than expecting any simplistic index to produce an informative outcome on its own.

A review of the literature reveals that measuring IDR has typically been conducted through either field-based research and surveys (Sanz-Menéndez, Bordons, & Zulueta, 2001; Palmer, 1999; Qin, Lancaster, & Allen, 1997) or through quantitative measures within a bibliometric approach or social network analysis (Schummer, 2004). Wagner et al. (2011) provide a full review of studies on the different approaches to understanding and measuring IDR, finding that bibliometric measures, such as co-authorships, co-inventors, collaborations, references, citations, and co-citations, are the most frequently studied and used. It is worth noting that bibliometric methods are not capable of discriminating among multidisciplinarity, interdisciplinarity, and transdisciplinarity. Bibliometric analyses can demonstrate some form of knowledge integration, but in no way do they reveal the modality and level of disciplinary integration.

Bibliometric approaches take publications as the subject of study and measure IDR in terms of the co-occurrence of discipline-specific elements, such as keywords, classification headings, the publishing journals, the authors, or the publications' reference lists. The most diffuse of these approaches is certainly citation analysis, where citations to papers in other disciplines are considered to be a signal of possible interactions or integration between different fields. Many studies have taken this bottom-up approach, building their case for IDR by measuring individual articles (Porter, Cohen, Roessner, & Perreault, 2007; Rafols & Meyer, 2010; Wang, Thijs, & Glänzel, 2015; Mugabushaka, Kyriakou, & Papazoglou, 2016; Zhang, Rousseau, & Glänzel, 2016). However, Zhang, Sun, Chinchilla-Rodríguez et al., (2018) point out that the concept of IDR is abstract and complex, which makes it difficult to identify a single indicator that can be used to fully represent or measure IDR.

A potential solution may be to conceive IDR in terms of the diversity of the disciplines involved. Rao (1982) discusses two general methods of obtaining measures of diversity within a population: One based on an intrinsic notion of dissimilarity between individuals and the other based on the concepts of entropy. According to Stirling (1994), "diversity" has three distinct components: "variety", "balance", and "disparity". Further, Stirling (2007) proposes suitable measurement indicators for each component. The variety indicator answers the question: "How many types of thing do we have?" For "balance", the question is: "How much of each type of thing do we have?" Finally, for "disparity" the question is: "How different from each other are the types of thing that we have?"

In the bibliometric sphere, these concepts have been widely applied to investigate IDR, as demonstrated in Wagner et al.'s review (2011). Porter and Rafols (2009) proposed measuring Stirling's (1994) three basic properties of research diversity by mapping the subject categories of cited publications. The authors introduced an "integration score", which indexes the number of disciplines cited by a paper along with their "concentration" and "cognitive distance". Rafols and Meyer (2010) maintain that the most appropriate indicator for studying the interdisciplinarity of a paper is the proportion of citations to papers in other disciplines. Recently, Zhang, Rousseau, and Glänzel (2016) proposed a new indicator, $^2D^3$, which is a monotone transformation of the Rao-Stirling indicator of diversity (Rao, 1982; Stirling, 2007). In an evaluation of physics research programs operating in the Netherlands in 1996, Rinia, van Leeuwen, and van Raan (2002) defined a measure of IDR programs as the percentage of articles originating from such programs that are published in journals attributed to other disciplines.

Each of these works are based on the assumption that an article reflects the same discipline as the publishing journal. In fact, this assumption is true of many other works on IDR (Levitt & Thelwall, 2008; Porter, Roessner, & Heberger, 2008; Adams, Jackson, & Marshall, 2007; Morillo, Bordons, & Gómez, 2003; Morillo, Bordons, & Gomez, 2001). Very few works identify IDR from the bibliometric perspective of co-authorship. The disciplinary field of an author in fact can be thought of as their knowledge contribution to the project. Although there are likely to be some differences between nations, disciplines, scientific communities, and single organizations, co-authorship is assumed to be a reliable indication of the contribution each scientist has made to the success of a specific research project. In particular, the disciplinary field of an author can be thought of as their knowledge contribution to the project. As Schummer (2004) states, "Co-author analysis measures IDR in terms of successful research interaction between disciplines". This notion shifts the problem of recognizing IDR through the semantic analysis of an article or the scientific classifications of the papers cited to that of identifying the specializations of its authors. In his investigations of patterns and degrees of interdisciplinarity in nanoscience research, Schummer (2004) applied this approach to a dataset of 600 papers published in "nano" journals in the 2002–2003 biennium. His study was based on observations about the departmental affiliations of each co-author. However, in addition to its limited scope of only 600 papers, the study suffers from using the departmental affiliation as a proxy for disciplinary affiliations, which is of questionable validity. In reality, departments are organizational units that may comprise members from several different disciplines. Using an approach based on the field classification of authors, Abramo, D'Angelo, and Di Costa (2012) analyzed the degree of collaboration among scientists from different disciplines to identify the most frequent "combinations of knowledge" in research activity, drawing on the 2004-2008 Web of Science (WoS) publications by all Italian universities' professors working in the sciences. More recently, Abramo, D'Angelo, and Di Costa (2017a) showed that interdisciplinary research teams deliver higher gains to science.

For reasons that will be apparent in the next section, the Italian case is particularly suited to the bibliometric analysis of IDR by the authors method, and forms the case analysis for this paper.

## 2. Data and methods

### 2.1 Data

The application of the authors method to measure IDR requires identifying the specialization of a paper's authors. In the Italian academic system, professors must classify themselves in one, and only one, scientific field. These fields, called scientific disciplinary sectors (SDSs), 370 in all, are each grouped under one of 14 university disciplinary areas (UDAs).[3] With the exception of Norway, it seems that no other country classifies their academics by discipline, which makes the Italian case particularly appropriate for this kind of analysis. The bibliometric data used in this study is extracted from the Italian Observatory of Public Research, a database developed and maintained by the Italian authors of this paper, and derived under license from the WoS core collection. Beginning from the raw data of the WoS publications with affiliation Italy, and applying a complex algorithm to reconcile the author's affiliation and disambiguation of the true identity of the authors, each publication by Italian universities' academics is attributed to the university professor or professors that produced it (D'Angelo, Giuffrida, & Abramo,

---

[3] In Italy, all personnel enter the university system and progress their careers through national public examinations. These examinations are given per field (SDS) and are assessed by members of the same SDS. Candidates must choose the SDS in which to compete and show their competence in that specific SDS through their research outputs. SDS classifications are governed by a very large committee of university professors across all scientific disciplines, called the CUN.

2010).[4]

In this work, we observe the 2006-2016 WoS indexed publications by Italian universities. The publication types are restricted to articles, proceedings papers, and book chapters since these types of documents normally contain original research, unlike reviews, letters, etc. In particular, reviews, sometimes called literature reviews, synthesize or overview research already conducted in primary sources. They generally summarize the current state of research on a given topic. Therefore the knowledge integration revealed from a "review" may carry a different cognitive meaning than the original article or proceeding papers (Zhang et al., 2012). Because only Italian academics are classified in the SDS system, we are forced to exclude also publications whose co-authors are affiliated to foreign institutions or Italian organizations other than universities. Finally, we exclude publications with no references indexed in the WoS (source items). The final dataset comprises 43,667 publications: 19,286 with single-authors and 24,381 with multiple authors.

**2.2 The disciplinary analysis of the reference list**

According to the reference list method, interdisciplinarity of a publication is measured by the diversity of the research fields represented in a publication's reference list. Operationally, we identify the fields of a referenced publication by the WoS subject categories (SCs) of the relevant journal (full counting). Diversity can be measured through the three components: variety, balance, and disparity, outlined in Stirling (2007). These three decompositions make it possible to explore different aspects of diversity in the cited references.

*Variety* is defined as the number of non-empty categories assigned to system elements. In this study, the system elements are the WoS-indexed references and the non-empty categories are the SCs in WoS. In formulae:

$$V = \sum_i SC_i \qquad [1]$$

Where $SC_i = 1$ if the ith SC is represented in the reference list; 0 otherwise. The value of V ranges between 1 and the number of SCs (currently 252).

*Balance* is a function of the pattern of the element assignments across categories – called "evenness" in ecology and "concentration" in economics. The Gini index is a well-known concentration measure where, if *G* denotes the Gini concentration measure, then B = 1-G is the corresponding measure of evenness or balance (Nijssen et al., 1998). In this study, we adopt "B = 1-G" as the balance indicator. In formulae:

$$B = 1 - \frac{\sum(2i-V-1)x_i}{V \sum x_i} \qquad [2]$$

Where *i* is the index ranging from 1 to *V*, $x_i$ is the number of references falling in the *i-th* SC, and the SCs are sorted by $x_i$ in non-decreasing order. The range of *B* is between 1/V (max concentration) and one (max balance).

---

[4] The harmonic mean of precision and recall (F-measure) of authorships disambiguated by the algorithm is around 95% (2% margin of error, 98% confidence interval). Additional manual disambiguation correcting false positive and negative authorships increases the harmonic mean to 98.5%.

***Disparity*** refers to the manner and degree to which things are distinguished. Disparity is the complement of similarity. In formulae,

$$\text{Dis}_{ij} = 1 - S_{ij} \quad [3]$$

Where $S_{ij}$ is the cosine similarity between the *i-th* SC and *j-th* SC, based on a cross-citation matrix for the period 1991-2015.[5] The range of $Dis_{ij}$ is between zero (max similarity) and one (max disparity).

In this study, we calculate the average disparity ($\overline{Dis}$) between the referenced SCs. In formulae:

$$\overline{\text{Dis}} = \frac{1}{V(V-1)} \sum_{i \neq j} \text{Dis}_{ij} \quad [4]$$

Additionally, we measure the Zhang, Rousseau, & Glänzel (2016) Integrated Diversity (ID) index, as the representative indicator of integrated diversity for the above three components.

***Integrated diversity*** is defined as,

$$\text{ID} = \frac{1}{\sum_{i \neq j} p_i p_j (1 - Dis_{ij})} \quad [5]$$

Where $p_i = {x_i}/{X}; X = \sum x_i$. In the case of one single SC, namely $\text{Dis}_{ij} = 0$, the value of integrated diversity, ID is 1.

**2.3 The disciplinary analysis of authors**

The SDS of a researcher is a reflection of their educational background, their expertise, and their primary field of research. However, this does not mean that their research is necessarily always confined to their SDS, in fact research diversification occurs indeed (Abramo, D'Angelo, & Di Costa, 2017b,c). For example, a statistician may join research teams in medicine, physics, social sciences, etc. giving rise to IDR. Once the true identity of each co-author is determined, a publication can then be associated with one or more SDSs, and its disciplinary nature assessed. In this work we apply the disciplinary analysis of authors to distinguish interdisciplinary publications from non-interdisciplinary ones. According to the authors method, a single-author paper, by convention, cannot be interdisciplinary since the author belongs to one, and only one, SDS. Nor can a publication co-authored by academics belonging to the same SDS be interdisciplinary. Single-SDS publications can then be further subdivided into single-author and multi-author (whereby all co-authors belong to the same SDS) publications. For each subpopulation we then measure its publication diversity by the reference list method, calculating the above four indicators presented in subsection 2.2. Differences in diversity are expected across subpopulations.

In theory, one would expect to find nil diversity in single-SDS publications when IDR is measured with the alternative approach, i.e. the reference list method. In practice, however, there may be a low level of diversity that we call "physiological". The reasons for this are:
- Research fields generally present blurring boundaries and some overlapping domains.

---

[5] The cross-citation matrix of all SCs (1991-2015) was constructed by Lin Zhang based on an in-house database of the Centre for R&D Monitoring (ECOOM), Belgium.

- Research scientists often have a broader educational background than their specialization (e.g., a vascular surgeon is, first of all, a physician), which allows them to integrate knowledge from cognitively close fields. Additionally, they may have had opportunities to diversify their research during their career, say, by joining multidisciplinary teams, which has provided them with exposure to theories and methods from different disciplines.
- The SDS classification system (370 SDSs in all) differs from the WoS subject classification scheme (252 SCs), which we use to analyze diversity in reference lists.

For the above reasons, we expect that the "physiological" diversity of (non-interdisciplinary) papers co-authored by scientists belonging to the same SDS, will grow with the number of authors. For (interdisciplinary) papers co-authored by scientists belonging to different SDSs, we would expect that diversity in the reference list grows along with the number of SDSs reflected in the byline. To a lesser extent, we would expect diversity to increase with the number of authors, being the number of SDSs equal. Furthermore, the disparity of SCs in the reference list should be reflective of the cognitive distance between the SDSs represented in the byline (i.e. when the SDSs fall under different UDAs).

## 2.4 Limits and assumptions

Before presenting the results of our analysis, we would like to warn the reader about the underlying assumptions of all bibliometric studies dealing with or based on discipline classification systems. In particular, IDR studies need refer to more or less agreed upon primary units of internal differentiation of science, called disciplines. The semantic roots of "*disciplina*" (discipline) are found in the Latin "*discere*" (learning), as a term for the ordering of knowledge for the purposes of education. Bibliographic repertories' (i.e. WoS and Scopus) discipline classification systems respond to the later archival function of disciplines: the discipline is a place where one deposits knowledge after having found it out, but it is not an active system for the production of knowledge. The problem is that one deposits new knowledge in an article, while WoS and Scopus index and classify journals, which may publish articles falling in different disciplines (hence, multi-category journals). IDR studies analyzing disciplinary diversity in the reference lists, and relying on journals' disciplinary classification systems, inevitably inherit the limits of associating a journal (and the articles therein) to one or more pre-defined disciplines.

Alternative disciplinary classification systems have been proposed in the literature. Waltman and Van Eck (2012) introduced a new methodology for constructing classification systems at the level of individual publications. In this approach, publications are clustered in research areas based on citation relations, and then labels are assigned based on extracting terms from the titles and abstracts. Unlike the WoS and Elsevier's ASJC (used in Scopus) systems, each publication is assigned to a single cluster. Ruiz-Castillo and Waltman (2015) compare the WoS with the various classification systems described above. Building a sequence of twelve classification systems, they focus on the consequences of adopting different granularity levels with increasing numbers of clusters. The analysis suggests that working with a few thousand significant clusters may be an optimal choice, and that algorithmically constructed systems can offer an up-to-date representation of the structure of scientific fields. Klavans and Boyack (2017) compare algorithmically-constructed publication-level classification systems, based on direct citations, to seven different journal classification schemes: ASJC; UCSD journal classification (Börner et al., 2012); Science-Metrix journal classification (Archambault, Beauchesne, & Caruso, 2011); Australian Research Council journal classification; KU Leuven ECOOM journal classification (Glanzel & Schubert, 2003); WoS SCs; United States National Science Foundation journal classification, used in the biannual Science & Engineering Indicators reports. They find that the document-based taxonomies provide a more accurate representation of disciplines than journal-based taxonomies. Most recently, Perianes-Rodríguez and Ruiz-Castillo (2017) propose a new criterion for choosing between journal-level and publication-level algorithmically-constructed classification systems (based on WoS), and recommend the second option.[6]

Furthermore, the reference list method assumes that a reference from another discipline directly translates into knowledge integration. While this could be generally the case, a number of exceptions may occur.[7]

The disciplinary classification of authors avoids the above traps. It reflects the rise of disciplines intended as production and communication systems, along with the specialization of scientists. Specialization is first of all an intellectual orientation. It depends on a decision to concentrate on a relatively small field of scientific activity. The emergence of communities of specialists is a further relevant and concomitant circumstance. In this respect the rise of disciplines is synonymous with the emergence of scientific communities theorized about since Thomas Kuhn (Kuhn, 1970). One of the limits of the above approach to study IDR, is represented by the availability of such classification schemes of scientists, in countries other than Italy and Norway (see subsection 2.1). Another limit is represented by the difficulty of classifying scientists who tend to significantly diversify their research.

A distinctive characteristic of the modern system of scientific disciplines is the dynamism resulting from i) the intensification of the interactions between ever more disciplines, which IDR studies aim at investigating; and ii) the expansionary strategy of a growing number of scientific communities, in which the discipline attacks and takes over parts of the domain of other disciplines. Such dynamism implies though ever changing disciplinary boundaries, which makes fine-grained classification schemes rapidly obsolete.

## 3. Analysis and results

We divide the entire dataset of 43,667 publications into three subpopulations:
   a) single-author (single-SDS) papers – 19,286 publications;[8]

---

[6] The most sophisticated publication-level algorithmically-constructed classification system is probably the CWTS clustering. There are a number of reasons why we avoid using it in this work. First, it is not public. Second, it embeds over 2,000 clusters, which would make practically all publications highly interdisciplinary by the reference list method. Third, it would entail comparing two essentially different classification systems with different levels and sizes (SDSs are 370 only).

[7] The argument here is similar to whether one believes that the norm is that scientists cite papers to recognize their influence, being aware that exceptions (uncitedness, undercitation, and overcitation) occur (Mertonian or normative theory of citing), or that the opposite is true, i.e. that citing to give credit is the exception, while persuasion is the major motivation for citing (social constructivism).

[8] The high share of single-author papers is due to the fact that we have restricted our analysis to publications authored by professors of Italian universities only.

b) multi-author, single-SDS papers – 16,624 publications; and
c) multi-SDS (multi-author) papers – 7757 publications.

The three decomposed diversity indicators: variety, balance and disparity, in addition to the integrated diversity indicator, are calculated by the reference list method, for each individual publication in each of the three sub-populations.

## 3.1 Analysis of diversity of single-author papers

From our analysis of diversity in the reference lists of single-author papers (Table 1), we observe remarkable differences across UDAs. Biology shows the highest values of integrated diversity and variety, with an average of more than 11 SCs reflected in the references for each publication. Chemistry is second by the above indicators and first by disparity. The social sciences, arts, and humanities, and "Mathematics and Computer Science" show the lowest diversity values.

As previously mentioned, in the authors method, a single-author paper cannot be interdisciplinary. However, against our expectations, the reference lists of many single-author papers showed high values of IDR. Table 2 presents the top five single-author papers by ID. It is worth noting that the first three of these papers also have the highest ID values across the entire dataset, which means that, according to the reference list method, the most interdisciplinary papers in our dataset are produced by single authors. Table 2 also shows that one author appears twice: Valerio, L, from Ist Nazl Fis Nucl, Sez Bologna. The SDS classification for this author is "Physics for Earth and Atmospheric Sciences". In his two single-author papers, he draws on references from many different SCs categories, such as "Anatomy & Morphology", "Biochemistry & Molecular Biology", "Chemistry, multidisciplinary", "Engineering, civil", "Environmental Sciences", "Forestry", "Limnology", "Material Science, multidisciplinary", "Mathematics", "Mechanics", "Statistics & Probability", "Water Resources", several computer science, physics, and engineering subjects, and so on.

As argued by Bunderson and Sutcliffe (2002), the diversity of a team can be defined by the distribution of disciplines across team members or by the extent to which the individuals who comprise the team are themselves diverse. For instance, a team may be diverse because it comprises specialized researchers in different fields (interpersonal diversity) or because the researchers themselves are interdisciplinary (intrapersonal diversity) (Banal-Estanol et al., 2018; Wagner et al., 2011). The researchers listed in Table 2 may serve as good examples of "intrapersonal diversity". What remains hard to explain though is that few of their publications present a diversity higher than all multi-fields (multi-author) papers in the dataset.

*Table 1: Descriptive statistics of diversity by the reference list method in single-author papers*

| UDA | No. Papers | % of Papers | Av. Variety | Av. Balance | Av. Disparity | Av. ID |
|---|---|---|---|---|---|---|
| Biology | 695 | 3.60% | 11.016 | 0.496 | 0.953 | 4.442 |
| Chemistry | 714 | 3.70% | 10.629 | 0.485 | 0.957 | 4.276 |
| Agricultural and Veterinary Sciences | 347 | 1.80% | 7.908 | 0.519 | 0.890 | 3.796 |
| Civil Engineering and Architecture | 1069 | 5.54% | 6.776 | 0.595 | 0.902 | 3.743 |
| Medicine | 1046 | 5.42% | 7.953 | 0.473 | 0.910 | 3.481 |
| Industrial and Information Engineering | 2789 | 14.46% | 5.730 | 0.608 | 0.858 | 3.439 |
| Earth Sciences | 335 | 1.74% | 7.322 | 0.524 | 0.921 | 3.426 |
| Economics and Statistics | 2410 | 12.50% | 5.602 | 0.481 | 0.862 | 2.867 |
| Physics | 2206 | 11.44% | 5.534 | 0.495 | 0.861 | 2.729 |
| Political and Social Sciences | 719 | 3.73% | 4.127 | 0.509 | 0.752 | 2.694 |
| History, Philosophy, Pedagogy, and Psychology | 1645 | 8.53% | 4.337 | 0.459 | 0.667 | 2.660 |
| Mathematics and Computer Science | 3843 | 19.93% | 4.184 | 0.479 | 0.761 | 2.367 |
| Law | 340 | 1.76% | 2.932 | 0.525 | 0.710 | 2.146 |
| Ancient History, Philology, Literature, and Art | 1128 | 5.85% | 2.168 | 0.389 | 0.526 | 1.760 |
| Total | 19286 | 100% | 5.565 | 0.503 | 0.811 | 2.950 |

*Table 2: Single-author papers with the highest integrated diversity indicator*

| Author | UDA | Publication Information | ID | Variety |
|---|---|---|---|---|
| Lucarini, V | Physics | From symmetry breaking to Poisson point process in 2D Voronoi tessellations: the generic nature of hexagons, *Journal of Statistical Physics*, 2008, 130, 6, 1947-1062. | 18.781 | 31 |
| Bianciardi, G | Medicine | Differential Diagnosis: Shape and Function, Fractal Tools in the Pathology Lab, *Nonlinear Dynamics Psychology and Life Sciences*, 2015, 19(4): 437-464. | 17.229 | 37 |
| Fiori, S | Industrial & Information Engineering | Fast statistical regression in presence of a dominant independent variable, *Neural Computing & Applications*, 2013, 22, 7-8, 1367-1378. | 16.909 | 21 |
| Lucia, U | Industrial & Information Engineering | Transport processes in biological systems: Tumoral cells and human brain, *Physica A-statistical Mechanics and its Applications*, 2014, 393, 327-336 | 16.060 | 28 |
| Lucarini, V | Physics | Three-Dimensional Random Voronoi Tessellations: From Cubic Crystal Lattices to Poisson Point Processes, *Journal of Statistical Physics*, 2009,134,1, 185-206. | 15.403 | 32 |

### 3.2 Analysis of diversity of multi-author single-SDS papers

In our dataset, 16,624 publications have multiple authors belonging to the same SDS. Thus, a question arises as to whether and to what extent the number of authors may affect diversity as measured by the reference list method. Table 3 presents the average diversity values for the reference lists of these multi-author single-SDS papers. The results for the single-author papers are shown for ease of comparison. Table 3 shows a monotonic increasing trend in all diversity indicators with an increase in the number of authors. The only exception is the balance indicator for "five or more-authors" publications. These results are not surprising, as in general, the bigger the research team, the higher the chances of knowledge integration across disciplinary boundaries. However, the relationship between the number of authors and the level of diversity does vary across

UDAs. The increasing trend is remarkable in "Economics and Statistics", "History, Philosophy, Pedagogy, and Psychology" and "Mathematics and Computer Science" which show an obvious increase in integrated diversity as the number of co-authors increases, while, in contrast, "Agricultural and Veterinary Sciences", "Civil Engineering and Architecture" and "Industrial and Information Engineering" show a relatively lower increase in integrated diversity as the number of co-authors grows.

*Table 3: Average diversity values (the reference list method) of single-SDS papers*

| No. Authors | No. Papers | Av. Variety | Av. Balance | Av. Disparity | Av. ID |
|---|---|---|---|---|---|
| 1 | 19286 | 5.565 | 0.503 | 0.811 | 2.950 |
| 2 | 10820 | 6.375 | 0.530 | 0.868 | 3.268 |
| 3 | 4344 | 6.499 | 0.550 | 0.882 | 3.358 |
| 4 | 1116 | 7.008 | 0.554 | 0.905 | 3.516 |
| 5 or more | 344 | 9.276 | 0.492 | 0.928 | 3.985 |

Table 4 presents the average values for each of the diversity indicators for the reference lists in multi-author single-SDS papers for each UDA. Similar to the results in Table 1, we found obvious differences in the levels of diversity across UDAs. Compared to Table 1, "History, Philosophy, Pedagogy, and Psychology" shows a dramatic change of position in Table 4 (from 11[th] to 2[nd] when ranked by average ID). Additionally, the distribution of publications is more concentrated than with single-author papers (Table 1), where more than half of the papers are classified into "Industrial and Information Engineering" and "Mathematics and Computer Science".

The up arrows in Table 4 indicate that the diversity of multi-author papers is higher than single-author (Table 1); the down arrows show the opposite. We observe that in most UDAs, multi-author papers have higher average diversity values than single-author papers in the same UDA (except for the balance indicator). These results indicate that, generally, multi-author papers cite publications falling in a wider range of SCs than single-author papers; moreover the cognitive distance (disparity) between such SCs is larger, which leads to higher integrated diversity values (ID). The only exceptions are found in "Chemistry", "Earth Sciences", "Industrial and Information Engineering", "Ancient History, Philology, Literature, and Art".

*Table 4: Average diversity values (the reference list method) of multi-author single-SDS papers*

| UDA | No. Papers | % of Papers | Avg. Variety | Av. Balance | Av. Disparity | Av. ID |
|---|---|---|---|---|---|---|
| Biology | 587 | 3.53 | 12.354 ↑ | 0.472 ↓ | 0.954 ↑ | 4.668 ↑ |
| History, Philosophy, Pedagogy, Psychology | 405 | 2.44 | 8.995 ↑ | 0.528 ↑ | 0.927 ↑ | 4.317 ↑ |
| Chemistry | 1134 | 6.82 | 10.779 ↑ | 0.458 ↓ | 0.956 ↓ | 4.146 ↓ |
| Agricultural and Veterinary Sciences | 701 | 4.22 | 8.738 ↑ | 0.512 ↓ | 0.944 ↑ | 3.983 ↑ |
| Civil Engineering and Architecture | 1550 | 9.32 | 6.914 ↑ | 0.614 ↑ | 0.932 ↑ | 3.887 ↑ |
| Medicine | 756 | 4.55 | 8.556 ↑ | 0.486 ↑ | 0.939 ↑ | 3.725 ↑ |
| Physics | 415 | 2.50 | 7.923 ↑ | 0.510 ↑ | 0.937 ↑ | 3.642 ↑ |
| Earth Sciences | 271 | 1.63 | 7.247 ↓ | 0.509 ↓ | 0.941 ↑ | 3.317 ↓ |
| Industrial and Information Engineering | 4846 | 29.15 | 5.392 ↓ | 0.615 ↑ | 0.861 ↑ | 3.291 ↓ |
| Economics and Statistics | 1913 | 11.51 | 6.487 ↑ | 0.503 ↑ | 0.903 ↑ | 3.242 ↑ |
| Political and Social Sciences | 62 | 0.37 | 4.419 ↑ | 0.500 ↓ | 0.845 ↑ | 2.740 ↑ |
| Law | 13 | 0.08 | 4.308 ↑ | 0.584 ↑ | 0.821 ↑ | 2.698 ↑ |
| Mathematics and Computer Science | 3941 | 23.71 | 4.468 ↑ | 0.478 ↓ | 0.787 ↑ | 2.439 ↑ |
| Ancient History, Philology, Literature, Art | 30 | 0.18 | 2.067 ↓ | 0.256 ↓ | 0.383 ↓ | 1.497 ↓ |
| Total | 16624 | 100 | 6.510 ↑ | 0.536 ↑ | 0.875 ↑ | 3.323 ↑ |

### 3.3 Analysis of diversity of multi-author multi-SDS papers

This subsection focuses on the 7757 multi-SDS (multi-author) papers. Table 5 presents the average diversity values (reference list method) of these publications, as a function of the number of SDSs reflected in the byline. Over 85% of papers only span two SDSs. Leaving aside the results for the 23 papers with five or more SDSs, not statistically meaningful, all dimensions of diversity increase as the number of SDSs increases.

*Table 5: Average diversity values (reference list method) of multi-SDS papers, as a function of the number of SDSs*

| No. SDSs | No. Papers | % of Papers | Av. Variety | Av. Balance | Av. Disparity | Av. ID |
|---|---|---|---|---|---|---|
| 2 | 6629 | 85.46% | 7.572 | 0.538 | 0.906 | 3.735 |
| 3 | 970 | 12.50% | 9.292 | 0.547 | 0.936 | 4.458 |
| 4 | 135 | 1.74% | 10.985 | 0.572 | 0.962 | 5.205 |
| 5 or more | 23 | 0.30% | 12.478 | 0.516 | 0.930 | 5.708 |

As further shown in Table 6,[9] average diversity generally increases with the number of authors, the number of SDSs being equal. The phenomenon is even more noticeable with the number of SDSs, the number of authors being equal. The balance indicator is, again, the exception with a peak at three authors for both two and three SDSs.

---

[9] We do not show the cases of more than three SDSs, because of the low number of observations.

*Table 6: Average diversity values (reference list method) of multi-SDS papers, as a function of the number of authors*

| No. Authors | No. Papers | % of Papers | Av. Variety | Av. Balance | Av. Disparity | Av. ID |
|---|---|---|---|---|---|---|
| No. SDSs=2 | | | | | | |
| 2 | 3173 | 47.87% | 7.215 | 0.529 | 0.895 | 3.580 |
| 3 | 2090 | 31.53% | 7.525 | 0.552 | 0.909 | 3.764 |
| 4 | 898 | 13.55% | 8.120 | 0.546 | 0.919 | 3.967 |
| 5 | 317 | 4.78% | 8.666 | 0.532 | 0.932 | 4.133 |
| 6 | 109 | 1.64% | 10.128 | 0.509 | 0.959 | 4.404 |
| No. SDSs=3 | | | | | | |
| 3 | 409 | 42.16% | 8.643 | 0.559 | 0.933 | 4.303 |
| 4 | 290 | 29.90% | 9.152 | 0.541 | 0.934 | 4.405 |
| 5 | 162 | 16.70% | 10.136 | 0.542 | 0.940 | 4.661 |
| 6 | 75 | 7.73% | 10.333 | 0.534 | 0.944 | 4.642 |

### 3.4 The distributions of integrated diversity

In this subsection we explore the distributions of ID across the three subpopulations of papers. The same exercise can be conducted for each dimension of diversity. Figure 1 shows the distributions ID of single-author papers, multi-author single-SDS papers, and multi-SDS papers. Single-author papers present the highest share of papers with the lowest diversity (ID = 1), with 13.7% of the papers reflecting one SC only in the cited references. For multi-author single-SDS papers the share drops to 7.1%; and for multi-SDS papers to 4.9%. Additionally, ID of single-author papers charts distinctly above the other two lines at the lower end of the diversity scale (to the left in Figure 1), but at greater levels of ID, above 3, it is multi-SDS papers that account for the highest proportion of papers.

Figure 2 shows a similar trend: in general, papers with bylines reflecting three SDSs account for a distinctly larger proportion of papers with higher ID than those with two SDSs, notably at the high end. However, although we find a general increase in the diversity of cited references from single authors to multi-author single-SDS authors, and finally, to multi-SDS authors, it should be noted that can be observed a range of diversity values in each subpopulation. There are papers published by single authors and multiple authors with a single SDS with extremely high diversity values, while a number of the multi-SDS papers present very low diversity (ID=1).

A closer examination of multi-SDS papers by UDA is presented in Table 7. It shows that the disparity is generally higher, when the SDSs reflected in the byline belong to different UDAs. This finding is as expected and indicates that, in general, the disparity measured by the reference list method reflects the cognitive distance between the SDSs in the byline. In general the average values of variety and ID increase as the number of UDAs increase. However, we note again that exceptions occur, as a few papers with more than three UDAs reflected in the byline, do show very low reference diversity.

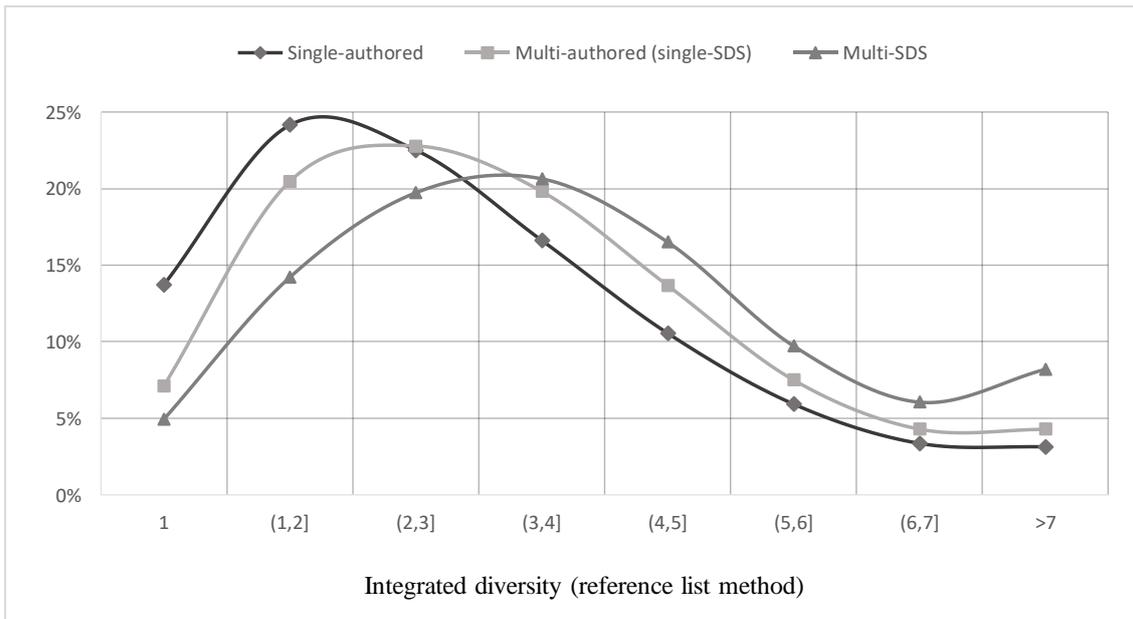

*Figure 1: Distributions of the integrated diversity (reference list method) in single-author, multi-author single-SDS, and multi-SDS papers*

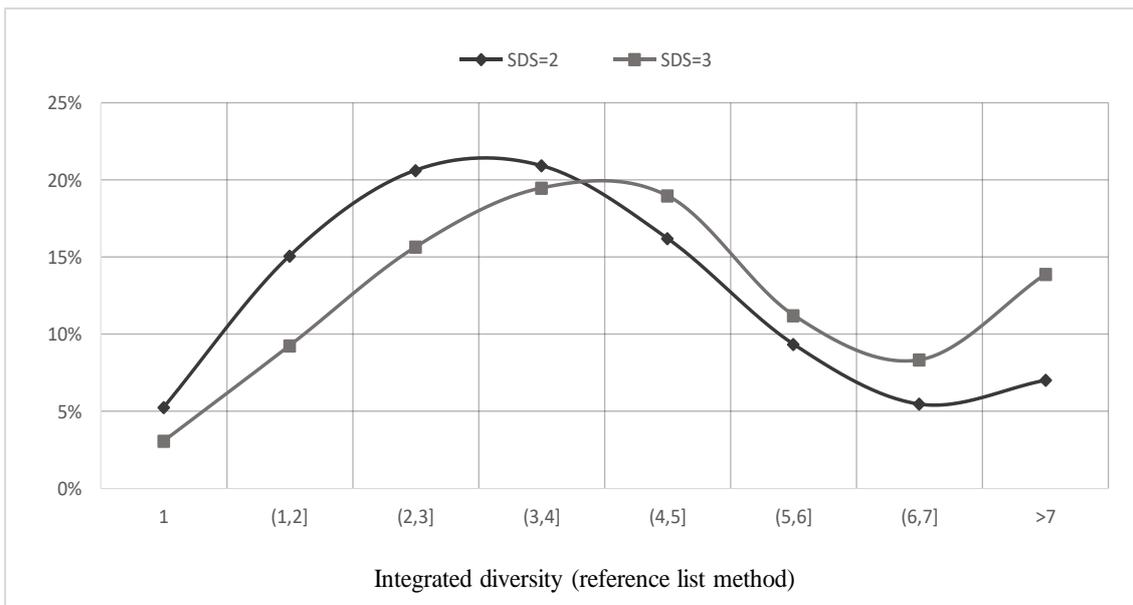

*Figure 2: Distributions of the integrated diversity (reference list method) in 2-SDS and 3-SDS papers*

*Table 7: Average diversity values (reference list method) of multi-SDS papers by number of UDAs*

| No. UDA | No. Papers | % of Papers | Av. Variety | Av. Disparity | Av. Balance | Av. ID |
|---|---|---|---|---|---|---|
| 1 | 4636 | 59.77% | 7.334 | 0.903 | 0.529 | 3.620 |
| 2 | 2960 | 38.16% | 8.545 | 0.921 | 0.556 | 4.171 |
| 3 or more | 161 | 2.08% | 10.466 | 0.947 | 0.549 | 4.897 |

Finally, Table 8 shows the descriptive statistics of the diversity indicators distributions for each subpopulation.

*Table 8: Descriptive statistics of diversity by the reference list method for papers in the three subpopulations*

|          | Variety |         |       | Balance |         |       | Disparity |         |       | Integrated Diversity |         |       |
|----------|---------|---------|-------|---------|---------|-------|-----------|---------|-------|----------------------|---------|-------|
|          | S-A     | M-A_S-SDS | M-SDS | S-A   | M-A_S-SDS | M-SDS | S-A     | M-A_S-SDS | M-SDS | S-A                | M-A_S-SDS | M-SDS |
| Average  | 5.565   | 6.510   | 7.860 | 0.503   | 0.536   | 0.540 | 0.811     | 0.875   | 0.911 | 2.950                | 3.323   | 3.857 |
| Median   | 4.000   | 5.000   | 7.000 | 0.512   | 0.533   | 0.536 | 0.958     | 0.963   | 0.969 | 2.544                | 2.987   | 3.526 |
| St. dev. | 4.490   | 4.640   | 5.048 | 0.285   | 0.248   | 0.223 | 0.331     | 0.252   | 0.212 | 1.773                | 1.819   | 2.028 |
| Min      | 1.000   | 1.000   | 1.000 | 0.000   | 0.000   | 0.000 | 0.000     | 0.000   | 0.000 | 1.000                | 1.000   | 1.000 |
| Max      | 60.00   | 51.00   | 39.00 | 1.000   | 1.000   | 1.000 | 1.000     | 1.000   | 1.000 | 18.78                | 16.56   | 14.60 |

S-A = single-author (single-SDS) papers
M-A_S-SDS = multi-author, single-SDS papers
M-SDS = multi-SDS (multi-author) papers

## 4. Discussion and Conclusions

As the advantages of IDR become more evident, policy initiatives to foster it increase (Rijnsoever and Hessels, 2011). There exist numerous barriers to collaboration among scientists belonging to different disciplines: i) epistemic barriers involving styles of thought; ii) research traditions, techniques, and language that are difficult to translate across disciplinary domains; iii) disciplinary structures involving specialized journals, conferences and academic societies; iv) administrative barriers (Jacobs & Frickel, 2009).

Jacobs and Frickel (2009) also raise the question of whether there is a "citation penalty" for IDR, meaning that to pursue interdisciplinary types of projects, the scholars would pay a price in terms of lack of recognition for their work. Levitt and Thelwall (2008) speak of disadvantage for interdisciplinary studies in the natural sciences, but not in the social sciences, while Rinia, van Leeuwen, and van Raan (2002) report no penalties to interdisciplinary publications. Leahey, Beckman, and Stanko (2017) have analyzed the tradeoff between penalties and benefits for interdisciplinary work and conclude that IDR is a high-risk, high-reward endeavor. However, it seems immature to evaluate the impact of IDR and of the policy initiatives to foster it, before metrics can adequately reflect and interpret the multi-faceted concept as interdisciplinarity.

Our study falls in the realm of works responding to the need for accurate measurements of IDR. In this study, we compared two approaches to measure interdisciplinary research outputs based on data from the Italian Observatory on Public Research. One approach analyses the disciplinary diversity of the authors of a publication; the other the disciplinary diversity of the references. We have divided the original publication dataset into three subpopulations, according to the diversity of the authors, namely single-author papers, multi-author single-SDS, and multi-SDS. We have then applied the reference list method to each subpopulation, and measured the three main dimensions of diversity, namely variety, balance, and disparity, and an integrated indicator of diversity.

We found that, in general, the disciplinary diversity of the reference list grows with the number of SDSs reflected in a paper's byline and, to a lesser extent, with the number of authors, the number of SDSs being equal. Further, we found that the higher the number of SDSs in the byline falling in different UDAs, the higher the disparity as measured by the reference list method. In spite of this general convergence, we also found numerous cases of publications which show contrasting results. One example for all, the case of a single-author publication with the highest reference diversity in the dataset. Cases of multi-SDS papers showing low reference diversity are not rare either. While a physiological level of interdisciplinarity may be found also in single-author publications, an excess of it as measured by the reference list method, which is common to a high number of publications, reveals anomalies which prompt future investigation. It is the intention of the authors to delve into such anomalous cases, with the aim of finding possible explanations, and proposing solutions to possible limits of either or both methods to measure diversity of research output. Our study confirms that more research is needed on IDR measurement. Different proxy indicators and approaches may deliver different insights about the interdisciplinary nature of the research under study. A multi-perspective framework for measuring interdisciplinarity in any unit of research combined with expert reviews and content interpretations is probably necessary and more informative.

**Acknowledgments**

Lin Zhang would like to acknowledge the support from the National Natural Science Foundation of China (Grant No.71573085), the Innovation Talents of Science and Technology in HeNan Province (Grant No.16HASTIT038, 2015GGJS-108), and the Excellence Scholarship in Social Science in HeNan Province (No.2018-YXXZ-10).

**References**

Abramo, G., D'Angelo, C.A., & Di Costa, F. (2012). Identifying interdisciplinarity through the disciplinary classification of coauthors of scientific publications. *Journal of the American Society for Information Science and Technology*, *63*(11), 2206–2222.

Abramo, G., D'Angelo, C.A., Di Costa, F. (2017a). Do interdisciplinary research teams deliver higher gains to science? *Scientometrics*, 111(1), 317-336.

Abramo, G., D'Angelo, C.A., Di Costa, F. (2017b). Specialization vs diversification in research activities: the extent, intensity and relatedness of field diversification by individual scientists. *Scientometrics*, 112(3), 1403–1418.

Abramo, G., D'Angelo, C.A., Di Costa, F. (2017c). The effects of gender, age and academic rank on research diversification. *Scientometrics,* 114(2), 373–387.

Adams J., Jackson L., & Marshall, S. (2007). Bibliometric analysis of interdisciplinary research. *Report to the Higher Education Funding Council for England. Leeds, UK: Evidence Ltd*.

Adams, J., Loach, T., & Szomszor, M. (2016): Interdisciplinary Research: Methodologies for Identification and Assessment. *Digital Research Reports*, Digital Science, London

Archambault, É., Beauchesne, O.H., & Caruso, J. (2011). Towards a multilingual, comprehensive and open scientific journal ontology. In *Proceedings of the 13th international conference of the international society for scientometrics and informetrics* (pp. 66-77), (ISSI 2011), Durban, South Africa, July 4-8.

Banal-Estanol, A., Macho-Stadler. I., & Perez-Castrillo, D. (2018), Team diversity evaluation by research grant agencies: Funding the seeds of radical innovation in academia? In preprint.

Börner, K., Klavans, R., Patek, M., Zoss, A.M., Biberstine, J.R., Light, R.P., ... & Boyack, K. W. (2012). Design and update of a classification system: The UCSD map of science. *PLoS ONE*, 7(7), e39464.

Bunderson, J.S., & Sutcliffe K.M. (2002). Comparing alternative conceptualizations of functional diversity in management teams: process and performance effects, *Academy of management journal*, 45(5): 875-893.

Choi, B. C. K., & Pak, A. W. P. (2006). Multidisciplinarity, interdisciplinarity and transdisciplinarity in health research, services, education and policy: 1. Definitions, objectives, and evidence of effectiveness. *Clinical and Investigative Medicine. Medecine Clinique et Experimentale*, *29*(6), 351–64.

Committee on the Science of Team Science; Board on Behavioral, Cognitive, and Sensory Sciences; Division of Behavioral and Social Sciences and Education; National Research Council; Cooke NJ, Hilton ML, editors. Enhancing the Effectiveness of Team Science. Washington (DC): National Academies Press (US); 2015 Jul 15. Available from: https://www.ncbi.nlm.nih.gov/books/NBK310387/ doi: 10.17226/19007.

D'Angelo, C.A., Giuffrida, C.& Abramo, G. (2010). A heuristic approach to author name disambiguation in large-scale bibliometric databases. *Journal of the American Society for Information Science and Technology*, 62(2), 257–269.

Glänzel, W., & Schubert, A. (2003). A new classification scheme of science fields and subfields designed for scientometric evaluation purposes. *Scientometrics*, 56(3), 357-367.

Huutoniemi, K., Klein, J. T., Bruun, H., & Hukkinen, J. (2010). Analyzing interdisciplinarity: Typology and indicators. *Research Policy*, *39*(1), 79–88.

Jacobs, J. A., & Frickel, S. (2009). Interdisciplinarity: A Critical Assessment. *Annual Review of Sociology*, *35*(1), 43–65.

Klavans, R., & Boyack, K.W. (2017). Which type of citation analysis generates the most accurate taxonomy of scientific and technical knowledge? *Journal of the Association for Information Science and Technology*, 68(4), 984-998.

Klein, J. T. (2008). Evaluation of interdisciplinary and transdisciplinary research. A literature review. *American Journal of Preventive Medicine*, *35*(2 Suppl), S116–S123.

Kuhn, T. S. (1970). *The Structure of Scientific Revolutions*, 2nd ed., University of Chicago Press, Chicago.

Leahey, E., Beckman, C. M., & Stanko, T. L. (2017). Prominent but Less Productive. *Administrative Science Quarterly*, *62*(1), 105–139.

Levitt, J. M., & Thelwall, M. (2008). Is multidisciplinary research more highly cited? A macrolevel study. *Journal of the American Society for Information Science and Technology*, *59*(12), 1973–1984.

Morillo, F., Bordons, M., & Gomez, I. (2001). An approach to interdisciplinarity through bibliometric indicators. *Scientometrics*, *51*(1), 203–222.

Morillo, F., Bordons, M., & Gómez, I. (2003). Interdisciplinarity in science: A tentative typology of disciplines and research areas. *Journal of the American Society for Information Science and Technology*, *54*(13).

Mugabushaka, A.-M., Kyriakou, A., & Papazoglou, T. (2016). Bibliometric indicators of interdisciplinarity: the potential of the Leinster–Cobbold diversity indices to study disciplinary diversity. *Scientometrics*, *107*(2), 593–607.

NAS/NAE/IOM. (2005). *Facilitating Interdisciplinary Research*. Washington, D.C.: National Academies Press.

National Science Board (2016). *Science and Engineering Indicators 2016*. Arlington, VA: National Science Foundation (NSB-2016-1).

Nijssen, D., Rousseau, R. and Van Hecke, P. (1998). The Lorenz curve: a graphical representation of evenness. *Coenoses*, *13*(1), 33-38.

Palmer, C. L. (1999). Structures and strategies of interdisciplinary science. *Journal of the Association for Information Science and Technology*, *50*(3), 242–253.

Perianes-Rodriguez, A., & Ruiz-Castillo, J. (2017). A comparison of the Web of Science and publication-level classification systems of science. *Journal of Informetrics*, 11(1), 32-45.

Porter, A. L. & Rafols, I. (2009). Is science becoming more interdisciplinary? Measuring and mapping six research fields over time. *Scientometrics, 81*(3), 719-745.

Porter, A. L., Cohen, A. S., Roessner, J., & Perreault, M. (2007). Measuring researcher interdisciplinarity. *Scientometrics*, *72*(1), 117–147.

Porter, A. L., Roessner, D. J., & Heberger, A. E. (2008). How interdisciplinary is a given body of research? *Research Evaluation*, *17*(4), 273–282.

Qin, J., Lancaster, F. W., & Allen, B. (1997). Types and levels of collaboration in interdisciplinary research in the sciences. *Journal of the American Society for Information Science*, *48*(10), 893–916.

Rafols, I., & Meyer, M. (2010). Diversity and network coherence as indicators of interdisciplinarity: Case studies in bionanoscience. *Scientometrics*, 82(2), 263–287.

Rao, C.R. (1982). Diversity: Its measurement, decomposition, apportionment and analysis. *Sankhya: The Indian Journal of Statistics*, Series A, 44(1), 1-22.

Rijnsoever, F. J. V., & Hessels, L. K. (2011). Factors associated with disciplinary and interdisciplinary research collaboration. *Research Policy, 40*(3), 463-472.

Rinia, E. J., van Leeuwen, T. N., & van Raan, A. F. J. (2002). Impact measures of interdisciplinary research in physics. *Scientometrics*, *53*(2), 241–248.

Rousseau, R., Zhang, L. & Hu, X.J. (2018). Knowledge Integration: its meaning and measurement. Book Chapter. *Springer Handbook of Science and Technology Indicators*, edited by W. Glänzel, H. Moed, U. Schmoch and M. Thelwall.

Ruiz-Castillo, J., & Waltman, L. (2015). Field-normalized citation impact indicators using algorithmically constructed classification systems of science. *Journal of Informetrics*, 9(1), 102-117.

Sanz-Menéndez, L., Bordons, M., & Zulueta, M. A. (2001). Interdisciplinarity as a multidimensional concept: its measure in three different research areas. *Research Evaluation*, *10*(1), 47–58.

Schummer, J. (2004). Multidisciplinarity, interdisciplinarity, and patterns of research collaboration in nanoscience and nanotechnology. *Scientometrics*, *59*(3), 425–465.

Stirling, A. (1994). Diversity and ignorance in electricity supply investment. *Energy Policy*, *22*(3), 195–216.

Stirling, A. (2007). A general framework for analysing diversity in science, technology and society. *Journal of The Royal Society Interface*, *4*(15), 707-719.

Stokols, D., Fuqua, J., Gress, J., Harvey, R., Phillips, K., Baezconde-Garbanati, L., … Trochim, W. (2003). Evaluating transdisciplinary science. *Nicotine & Tobacco Research*, *5*(6), 21–39.

U.S. National Research Council (2015). *Enhancing the Effectiveness of Team Science*. (Nancy J. Cooke and Margaret L. Hilton, Ed.). Washington, D.C.: National Academies Press.

Wagner, C. S., Roessner, J. D., Bobb, K., Klein, J. T., Boyack, K. W., Keyton, J., … Börner, K. (2011). Approaches to understanding and measuring interdisciplinary scientific research (IDR): A review of the literature. *Journal of Informetrics*, *5*(1), 14–26.

Waltman, L., & van Eck, N.J. (2012). A new methodology for constructing a publication-level classification system of science. *Journal of the American Society for Information Science and Technology*, 63(12), 2378-2392.

Wang, J., Thijs, B., & Glänzel, W. (2015). Interdisciplinarity and Impact: Distinct Effects of Variety, Balance, and Disparity. *PLOS ONE*, *10*(5).

Zhang, L., Glänzel, W. (2012). Proceeding papers in journals versus the regular journal publications. *Journal of Informetrics*, 6, 88-96.

Zhang, L., Rousseau, R., & Glänzel, W. (2016). Diversity of references as an indicator of the interdisciplinarity of journals: Taking similarity between subject fields into account. *Journal of the Association for Information Science and Technology*, *67*(5), 1257–1265.

Zhang, L., Sun, B., Chinchilla-Rodríguez, Z., Chen, LX., Huang, Y. (2018). Interdisciplinarity and collaboration: On the relationship between disciplinary diversity in departmental affiliations and reference lists. *Scientometrics*, DOI: 10.1007/s11192-018-2853-0.